\documentstyle[psfig,epsf]{article}
\begin{document}

\begin{center}
{\Large}{\bf LOOP QUANTUM GRAVITY EFFECTS ON THE HIGH ENERGY \\ COSMIC RAY SPECTRUM} \\

\vspace{4mm}

Jorge Alfaro \\
{\it Facultad de F\'{\i}sica, Pontificia Universidad Cat\'{o}lica de Chile \\
Casilla 306, Santiago 22,\\
Chile. }\\

\vspace{2mm}

and

\vspace{2mm}

Gonzalo A. Palma \\
{\it Department of Applied Mathematics and Theoretical Physics,\\
Center for Mathematical Sciences, University of Cambridge,\\
Wilberforce Road, Cambridge CB3 0WA,\\
United Kingdom.}
\end{center}

\begin{abstract}
Recent observations on ultra high energy cosmic rays (those cosmic
rays with energies greater than $\sim 4 \times 10^{18}$ eV)
suggest an abundant flux of incoming particles with energies above
$1 \times 10^{20}$ eV. These observations violate the
Greisen-Zatsepin-Kuzmin cutoff. To explain this anomaly we argue
that quantum-gravitational effects may be playing a decisive role
in the propagation of ultra high energy cosmic rays. We consider
the loop quantum gravity approach and provide useful techniques to
establish and analyze constraints on the loop quantum gravity
parameters arising from observational data. In particular, we
study the effects on the predicted spectrum for ultra high energy
cosmic rays and conclude that is possible to reconcile
observations.
\end{abstract}

\section{Introduction} \label{INTRO}

In the present discussion we are concerned with the observation of
ultra high energy cosmic rays (UHECR), i.e. those cosmic rays with
energies greater than $\sim 4 \times 10^{18}$ eV. Although not
completely clear, it has been suggested that these high energy
particles are possibly heavy nuclei \cite{exp1-Ave} (we will
assume here that they are protons) and, by virtue of the isotropic
distribution with which they arrive to us, that they originate in
extragalactic sources.

The propagation of UHECR in intergalactic space is subject to
interaction with the cosmic microwave background radiation (CMBR).
Its presence produces friction on UHECR making them release energy
in the form of secondary particles. This affects their possibility
to reach great distances. A first estimation of the characteristic
distance that UHECR can reach before losing most of their energy
was simultaneously made in 1966 by K. Greisen \cite{Greisen} and
G. T. Zatsepin \& V. A. Kuzmin \cite{Zatsepin & Kuzmin}. They
showed that the observation of cosmic rays with energies greater
than $4 \times 10^{19}$ eV should be greatly suppressed. This
energy is usually referred as to the GZK cutoff energy. A few
years later, F.W. Stecker \cite{Stecker} computed the mean life
time for protons as a function of their energy, giving a more
accurate perspective of the energy dependence of the cutoff and
showing that cosmic rays with energies above $1 \times 10^{20}$ eV
should not travel more than $\sim 100$ Mpc. More detailed
approaches to the GZK-cutoff feature have been made since these
first estimations. For example V. Berezinsky \& S.I. Grigorieva
\cite{Berezinsky2}, V. Berezinsky {\it et al.} \cite{Berezinsky}
and S.T. Scully \& F.W. Stecker \cite{Scully & Stecker} have made
progress in the theoretical study of the spectrum $J(E)$ (i.e. the
flux of arriving particles as a function of the observed energy
$E$) that UHECR should present. As a result, the GZK cutoff exists
in the form of a suppression in the predicted flux of cosmic rays
with energies above $\sim 8 \times 10^{19}$ eV.

Currently there are two main different sets of data for the
observed flux $J(E)$ in its most energetic sector ($E > 4 \times
10^{18}$ eV). On one hand we have the observations from the High
Resolution Fly's Eye (HiRes) collaboration group \cite{HiRes},
which seem to be consistent with the predicted theoretical
spectrum and, therefore, with the presence of the GZK cutoff.
Meanwhile, on the other hand, we have the observations from the
Akeno Giant Air Shower Array (AGASA) collaboration group
\cite{exp3}, which reveal an abundant flux of incoming cosmic rays
with energies above $1 \times 10^{20}$ eV. The appearance of these
high energy events is opposed to the predicted GZK cutoff, and
represents a great challenge that has motivated a vast amount of
new ideas and mechanisms to explain the phenomenon \cite{ideas}.
If the AGASA observations are correct then, since there are no
known active objects in our neighborhood (let us say within a
radius $R \simeq 100$ Mpc) able to act as sources of such
energetic particles and since their arrival is mostly isotropic
(without any privileged local source), we are forced to conclude
that these cosmic rays come from distances larger than $100$ Mpc.
This is commonly referred as the Greisen-Zatsepin-Kuzmin (GZK)
anomaly.

One of the interesting notions emerging from the possible
existence of the GZK anomaly is that, since ultra high energy
cosmic rays involve the highest energy events registered up to
now, then  a possible framework to understand and explain this
phenomena could be of a quantum-gravitational nature \cite{grav,
Alfaro & Palma 1, Alfaro & Palma 2}. This possibility is indeed
very exciting if we consider the present lack of empirical support
for the different approaches to the problem of gravity
quantization. In the context of the UHECR phenomena, all these
different approaches motivated by different quantum gravity
formulations, have usually converged on a common path to solve and
explain the GZK anomaly: the introduction of effective models for
the description of high energy particle propagation. These
effective models, pictures of the yet unknown full quantum gravity
theory, offer the possibility to modify conventional physics
through new terms in the equations of motion (now effective
equations of motion), leading to the eventual breakup of
fundamental symmetries such as Lorentz invariance (expected to be
preserved at the fundamental level). These Lorentz symmetry
breaking mechanisms are usually referred as Lorentz invariance
violations (LIV's) if the break introduce a privileged reference
frame, or Lorentz invariance deformations (LID's), if such a
reference frame is absent \cite{LID}. Its appearance on
theoretical and phenomenological grounds (such as high energy
astrophysical phenomena) has been widely studied, and offers a
large and rich amount of new signatures that deserve attention
\cite{LIV}.

To deepen the above ideas, we have adopted the loop quantum
gravity (LQG) theory \cite{LQG}. It is possible to study LQG
through effective theories that take into consideration
matter-gravity couplings. In this line, in the works of J. Alfaro
{\it et al}. \cite{AMU}, the effects of the loop structure of
space at the Planck level are treated semiclassically through a
coarse-grained approximation. An interesting feature of these
methods is the explicit appearance of the Plank scale $l_{p}$ and
the appearance of a new length scale ${\mathcal L} \gg l_{p}$
(called the ``weave'' scale), such that for distances $d \ll
{\mathcal L}$ the quantum loop structure of space is manifest,
while for distances $d \geq {\mathcal L}$ the continuous flat
geometry is regained. The presence of these two scales in the
effective theories has the consequence of introducing LIV's to the
dispersion relations $E=E(p)$ for particles with energy $E$ and
momentum $p$. It can be shown that these LIV's can significantly
modify the kinematical conditions for a reaction to take place.
For instance, as shown in detail in \cite{Coleman & Glashow}, a
consequence for the UHECR phenomenology is that the kinematical
conditions for a reaction between a primary cosmic ray and a CMBR
photon can be modified, leading to new effects and predictions
such as an abundant flux of cosmic rays well beyond the GZK cutoff
energy (explaining in this way the AGASA observations).

In this work we provide some techniques to establish and analyze
new constraints on the LQG parameters (or any LIV parameters
coming from other theories). Also, we attempt to predict a
modified UHECR spectrum consistent with the AGASA observations.
The results shown here are presented in detail in a previous work
by J. Alfaro and G. Palma \cite{Alfaro & Palma 2}. This paper is
organized as follows: In section \ref{UHECR} we give a brief
derivation of the conventional spectrum and analyze it with AGASA
observations. In section \ref{LQG} we present a short outline of
loop quantum gravity and its effective description of fermionic
and bosonic fields. In section \ref{Trh} we analyze the effects of
LQG corrections on the threshold conditions for the main reactions
involved in the UHECR phenomena to take place. In section
\ref{MOD-SPEC} we show how the modified kinematics can be relevant
to the theoretical spectrum $J(E)$ of cosmic rays (we will present
the obtained modified spectrum). In section \ref{CONC} we give
some final remarks.


\section{Ultra High Energy Cosmic Rays} \label{UHECR}

Here we review the main steps in the derivation of the UHECR
spectrum. This presentation will be useful and relevant to
understand how LQG corrections affect the predicted flux of cosmic
rays. The following material is mainly contained in the works of
F.W. Stecker \cite{Stecker}, Berezinsky {\it et al.}
\cite{Berezinsky} and S.T. Scully \& F.W. Stecker \cite{Scully &
Stecker}.

\subsection{General Description}

Two simple and common assumptions used in the development of the
cosmic ray spectrum are: 1) sources are uniformly distributed in
the Universe, and 2) the generation flux $F(E_{g})$ of emitted
cosmic rays from the sources is well described by a power law
behavior of the form $F(E_{g}) \propto E_{g}^{-\gamma_{g}}$, where
$E_{g}$ is the energy of the emitted particle and  $\gamma_{g}$ is
the generation index.

One of the main quantities for the computation of the UHECR
spectrum is the energy loss $- E^{-1} dE/dt$. This quantity
describes the rate at which a cosmic ray loses energy, and takes
into consideration two chief contributions: the energy loss due to
the redshift attenuation and the energy loss due to collisions
with the CMBR photons. This last contribution depends, at the same
time, on the cross sections $\sigma$ and the inelasticities $K$ of
the interactions produced during the propagation of protons in the
extragalactic medium, as well as on the CMBR spectrum. The most
important reactions taking place in the description of proton's
propagation are the pair creation and the photo-pion production:
\begin{equation}
p + \gamma \rightarrow p + e^{-} + e^{+}, \label{pair} \quad
\qquad \mathrm{and} \qquad \qquad  p + \gamma \rightarrow p + \pi.
\label{pion}
\end{equation}
The photo-pion production happens through several channels (for
example the baryonic $\Delta$ and $N$, and mesonic $\rho$ and
$\omega$ resonance channels, just to mention some of them) and is
the main responsible in the appearance of the GZK cutoff.

\subsection{Some Kinematics}

To study the interaction between protons and the CMBR, it is
useful to distinguish between three reference systems; the
laboratory system $\mathcal{K}$ (which we identify with the
Friedman Robertson Walker (FRW) co-moving reference system), the
center of mass (c.m.) system $\mathcal{K}^{*}$, and the system
where the proton is at rest $\mathcal{K}'$. In terms of these
systems, the photon energy will be expressed as $\omega$ in
$\mathcal{K}$ and as $\epsilon$ in $\mathcal{K}'$. The relation
between both quantities is simply $\epsilon = \gamma \omega
(1-\beta \cos \theta)$, where $\gamma = E/m_{p}$ is the Lorentz
factor relating $\mathcal{K}$ and $\mathcal{K}'$, $E$ and $m_{p}$
are the energy and mass of the incident proton, $\beta =
\sqrt{1-\gamma^{-2}}$, and $\theta$ is the angle between the
momenta of the photon and the proton measured in the laboratory
system $\mathcal{K}$. To determine the total energy
$E^{*}_{\mathrm{tot}} = E^{*} + \epsilon^{*}$ in the c.m. system,
it is enough to use the invariant energy squared $s \equiv
E_{\mathrm{tot}}^{2}-p_{\mathrm{tot}}^{2}$ (where
$E_{\mathrm{tot}} = E + \omega$ and $p_{\mathrm{tot}}$ are the
total energy and momentum in the laboratory system). In this way,
we have $E^{* \, 2}_{\mathrm{tot}} = s = m_{p}^{2} + 2 m_{p}
\epsilon$. As a consequence, the Lorentz factor $\gamma_{c}$ which
relates the $\mathcal{K}$ reference system with the
$\mathcal{K}^{*}$ system, is
\begin{eqnarray}
\gamma_{c}=\frac{E+\omega}{\sqrt{s}} \simeq \frac{E}{(m_{p}^{2} +
2 m_{p} \epsilon)^{1/2}}.
\end{eqnarray}

Let us consider the relevant case in which the reaction between
the proton and the CMBR photon is of the type
\begin{eqnarray}
p + \gamma \rightarrow a + b,
\end{eqnarray}
where $a$ and $b$ are two final particles of the collision. The
final energies of these particles are easily determined by the
conservation of energy-momentum. In the $\mathcal{K}^{*}$ system
these are
\begin{eqnarray}
E_{a, \, b}^{*}= \frac{1}{2 \sqrt{s}} (s + m_{a, \, b}^{2} - m_{b,
\, a}^{2}).
\end{eqnarray}
Transforming this quantity to the laboratory system, and averaging
with respect to the angle between the directions of the final
momenta, it is possible to find that the final average energy of
$a$ (or $b$) in the laboratory system is
\begin{eqnarray} \label{eq2: E a-b}
\langle E_{a, \, b} \rangle = \frac{E}{2} \left( 1+ \frac{m_{a, \,
b}^{2} - m_{b, \, a}^{2}}{s} \right).
\end{eqnarray}

The inelasticity $K$ of the reaction is defined as the average
fractional difference $K = \Delta E / E$, where $\Delta E = E -
E_{\mathrm{f}}$ is the difference between the initial energy $E$
and final energy $E_{\mathrm{f}}$ of the proton (in a single
collision with the CMBR photons). For the particular case of the
emission of an arbitrary particle $a$ (that is to say $p + \gamma
\rightarrow p + a$), expression (\ref{eq2: E a-b}) allows us to
write
\begin{eqnarray} \label{eq2: K}
K_{a}(s) = \frac{1}{2} \left( 1 + \frac{m_{a}^{2}-m_{p}^{2}}{s}
\right),
\end{eqnarray}
where $K_{a}$ is the inelasticity of the described process. This
is one of the main quantities involved in the study of the UHECR
spectrum.

\subsection{Mean Life $\tau (E)$}

To derive the UHECR spectrum it is imperative to know the mean
life $\tau (E)$ of the cosmic ray (or proton) with energy $E$
propagating in space, due to the attenuation of its energy by the
interactions with the CMBR photons. The mean life $\tau (E)$ is
defined through the relation $\tau (E)^{-1} = \left( -E^{-1} dE/dt
\right)_{\mathrm{col}}$, where the label ``col'' refers to the
fact that the energy loss is due to the collisions with the CMBR
photons. It is possible to show that the mean life $\tau (E)$ can
be written in the form
\begin{eqnarray} \label{eq2: tau5}
\tau (E)^{-1}= -\frac{kT}{2 \pi^{2} \gamma^{2}} \!
\int_{\epsilon_{\mathrm{th}}}^{\infty} \!\!\!\! d \epsilon \,
\sigma (\epsilon) K(\epsilon)  \epsilon  \ln [1-e^{-\epsilon/ 2
\gamma k T}],
\end{eqnarray}
where $K(\epsilon) = \Delta E / E$ is the inelasticity (with
$\Delta E$ the difference between the initial and final energies
of the proton before and after each collision), $\sigma(\epsilon)$
is the scattering cross section with the CMBR photons, $KT$ is the
temperature of the CMBR bath, $\gamma$ the lorentz factor, and
$\epsilon_{\mathrm{th}}$ is the threshold photon energy for
reactions to take place.

\subsection{Energy Loss and Spectrum}

The energy loss suffered by a very energetic proton during its
journey, from a distant source to our detectors, is not only
produced by the collisions that it has with CMBR at a particular
epoch. There is also a decrease in its energy due to the redshift
attenuation produced by the expansion of the Universe. At the same
time, such expansion will affect the collision rate through the
attenuation of the photon gas density, which can be understood as
a cooling of the CMBR through the relation $T = (1+z) T_{0}$,
where $z$ is the redshift and $T_{0}$ is the temperature of the
background at the present time. To calculate the spectrum we need
to consider the rate of energy loss during any epoch $z$ of the
Universe.

For the present discussion, we consider a flat Friedman Robertson
Walker (FRW) universe dominated by matter. The above assumptions
give rise to the following relation between the time coordinate
$t$ and the redshift $z$: $dt = - dz/H_{0} (1+z)^{5/2}$, where
$H_{0}$ is the Hubble constant at present time. Since the momentum
of a free particle in a FRW space behaves as $p \propto (1 + z)$,
we will have, with the additional consideration $p \gg m$ (where
$m$ is the particle mass), that the energy loss due to redshift is
\begin{eqnarray} \label{eq2: redshift contribution}
\left( -\frac{1}{E} \frac{dE}{dt} \right)_{\mathrm{cr}} = H_{0}
(1+z)^{3/2}.
\end{eqnarray}
On the other hand, the energy loss due to collisions with the CMBR
evolves as the background temperature changes (recall $T = (1+z)
T_{0}$). This evolution can be parameterized through $z$ and is
given by
\begin{eqnarray} \label{eq2: Evol of Gamma}
\left( -\frac{1}{E} \frac{dE}{dt} \right)_{\mathrm{col}} =
(1+z)^{3} \tau ([1+z]E)^{-1}.
\end{eqnarray}
Thus, the total energy loss can be expressed considering both
contributions (using $z$ instead of $t$):
\begin{eqnarray} \label{eq2: diff E-z}
\frac{1}{E} \frac{dE}{dz} = (1+z)^{-1} + H_{0}^{-1} (1+z)^{1/2}
\tau ([1+z]E)^{-1}.
\end{eqnarray}
Equation (\ref{eq2: diff E-z}) can be integrated numerically to
provide the energy $E_{g}(E,z)$ of a proton generated by the
source in a $z$ epoch and that will be detected with a energy $E$
here on Earth. Let us express this solution formally by:
$E_{g}(E,z)=\lambda(E,z) E$.

It is also possible to manipulate equation (\ref{eq2: diff E-z})
to obtain an expression for the dilatation of the energy interval
$dE_{g}/dE$. To accomplish this it is necessary to integrate
(\ref{eq2: diff E-z}) with respect to $z$ and then differentiate
it with respect to $E$ to obtain an integral equation for $d E_{g}
/ dE$. The solution of such an equation is found to be
\begin{eqnarray} \label{eq: dE-dE}
\frac{dE_{g}(z_{g})}{dE} = (1+z_{g}) \exp \left[ \int^{z_{g}}_{0}
\frac{dz}{H_{0}} (1+z)^{1/2} \frac{db(E')}{dE'} \right].
\end{eqnarray}
where $E'=(1+z)\lambda (E,z) E$.

The total flux $dj(E)$ of emitted particles from a volume element
$dV=R^{3}(z)r^{2} dr d \Omega$, in the epoch $z$ and coordinate
$r$, measured from  Earth at present time with energy $E$ is
\begin{eqnarray} \label{eq2: dj}
dj(E)dE = \frac{F(E_{0},z) dE_{0} n(z) dV}{(1+z) 4 \pi R_{0}^{2}
r^{2}},
\end{eqnarray}
where $j(E)$ is the particle flux per energy, $F(E_{0},z) dE_{0}$
the emitted particle flux within the range $(E_{0}, E_{0} +
dE_{0})$, and $n(z)$ the density of sources in $z$. As previously
mentioned, it is convenient to study the emission flux with a
power law spectrum of the type $F(E) \propto E^{-\gamma_{g}}$. It
can be shown that with such assumption, the relation between the
emission flux and the total luminosity $L_{p}$ of the source is
$F(E) = (\gamma_{g} - 2) L_{p} E^{-\gamma_{g}}$. To describe the
evolution of the sources we shall also use a power law behavior.
This will be done through the relation
\begin{eqnarray}
L_{p}(z)n(z) = (1+z)^{(3+m)} {\mathcal L}_{0},
\end{eqnarray}
where ${\mathcal L}_{0} = L_{p}(0)n(0)$. In this way, $m=0$
corresponds to the case in which sources do not evolve. If we
consider that $R_{0} = (1+z) R(z)$ and $R(z)dr=dt$ for flat spaces
(and $v \simeq 1$ for very energetic particles) and integrating
(\ref{eq2: dj}) from $z=0$ to some $z=z_{max}$ for which sources
are not relevant for the phenomena, it is possible to obtain
\begin{eqnarray} \label{eq2: j}
j(E) = (\gamma_{g}-2) \frac{1}{4 \pi} \frac{{\mathcal
L}_{0}}{H_{0}} E^{-\gamma_{g}}   \int^{z_{\mathrm{max}}}_{0}
\!\!\!\!\!\! dz_{g} (1+z_{g})^{m-5/2}
\lambda^{-\gamma_{g}}(E,z_{g}) \frac{dE_{g}(z_{g})}{dE}.
\end{eqnarray}
The above expression constitutes the spectrum of UHECR. The
volumetric luminosity ${\mathcal L}_{0}$ and the $\gamma_{g}$ and
$m$ indexes are free parameters that must be fixed
observationally.

\subsection{Ultra High Energy Cosmic Rays Spectrum}

To accomplish the computation of the theoretical spectrum we need
information about the dynamical processes taking place in the
propagation of protons along the CMBR. As we already emphasized,
the most important reactions taking place in the description of a
proton's propagation are, the pair creation $ p + \gamma
\rightarrow p + e^{-} + e^{+}$, and the photo-pion production $p +
\gamma \rightarrow p + \pi$. This last reaction is mediated by
several channels. The main channels are $N + \pi$, $\Delta + \pi$,
$R $, $N + \rho(770)$, $N + \omega(782)$. The total cross sections
and inelasticities of these processes are well known and can be
used in (\ref{eq2: tau5}) to compute the main time life of protons
as a function of their energy. Then, with the help of expressions
(\ref{eq: dE-dE}) and (\ref{eq2: j}), we can finally find the
predicted spectrum for UHECR.
\begin{figure}[tp] 
\centerline{\psfig{figure=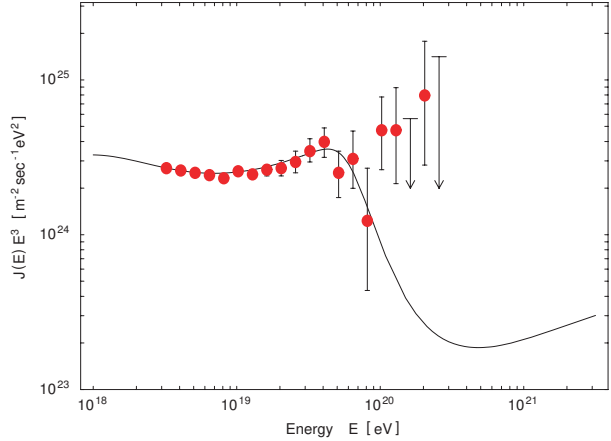,width=8cm}}
\caption{\label{F1} UHECR spectrum and AGASA observations. The
figure shows the UHECR spectrum $J(E)$ multiplied by $E^{3}$, for
uniform distributed sources, without evolution, and with a maximum
generation energy $E_{\mathrm{max}} = \infty$. Also shown are the
AGASA observed events. The best fit for the low energy sector ($E
< 4 \times 10^{19}$ eV) corresponds to $\gamma_{g} = 2.7$. }
\end{figure}
Fig.\ref{F1} shows the obtained spectrum $J(E)$ of UHECR and the
AGASA observed data. Again, we have selected for the theoretical
spectrum $J(E)$ the idealized case $E_{\mathrm{max}} = \infty$. To
reconcile the data of the low energy region ($E < 4 \times
10^{19}$ eV), where the pair creation dominates the energy loss,
it is necessary to have a generation index $\gamma_{g} = 2.7$
(with the additional supposition that sources do not evolve) and a
volumetric luminosity ${\mathcal L}_{0} = 4.7 \times 10^{51}
\mathrm{ergs}/\mathrm{Mpc}^{3}\mathrm{yr}$. It can be seen that
for events with energies $E > 4 \times 10^{19}$ eV, where the
energy loss is dominated by the photo-pion production, the
predicted spectrum does not fit the data well. To have a
statistical sense of the discrepancy between observation and
theory, we can calculate the Poisson probability $P$ of an excess
in the five highest energy bins. This is $P = 1.1 \times 10^{-8}$.
Another statistical measure is provided by the Poisson $\chi^{2}$
\cite{PDG}. Computing this quantity for the eight highest energy
bins, we obtain $\chi^{2} = 29$. These quantities show how far the
AGASA measurements are from the theoretical prediction given by
the curve of Fig.\ref{F1}. Other more sophisticated models have
also been analyzed in detail \cite{Berezinsky}, nevertheless, it
has turned out that the conventional standard model of physics
does not have the capacity to reproduce the observations from the
AGASA collaboration group in a satisfactory way.


\section{Loop Quantum Gravity} \label{LQG}

Loop quantum gravity is a canonical approach to the problem of
gravity quantization. It is based on the construction of a spin
network basis, labelled by graphs embedded in a three dimensional
insertion $\Sigma$ in space-time. A consequence of this approach
is that the quantum structure of space-time will be of a
polymer-like nature, highly manifested in phenomena involving the
Planck scale $l_{p}$.

The former brief description of loop quantum gravity allows us to
realize how complicated a full treatment of a physical phenomena
could be when the quantum nature of gravity is considered, even
for a flat geometry. It is possible, however, to introduce a loop
state which approximates a flat 3-metric on $\Sigma$ at length
scales greater than the length scale ${\mathcal L} \gg l_{p}$. For
pure gravity, this state is referred to as the weave state $|W
\rangle$, and the length scale ${\mathcal L}$ as the weave scale.
A flat weave $|W \rangle$ will be characterized by ${\mathcal L}$
in such a way that for distances $d \ll {\mathcal L}$ the quantum
loop structure of space is manifest, while for distances $d \geq
{\mathcal L}$ the continuous flat geometry is regained. With this
approach, for instance, the metric operator $\hat q_{a b}$
satisfies
\begin{eqnarray} \label{eq3: weave-g}
\langle W| \hat q_{a b} |W \rangle = \delta_{ab} + {\mathcal O}
(l_{p} / {\mathcal L}).
\end{eqnarray}

A generalization of the former idea, to include matter fields, is
also possible. In this case, the loop state represents a matter
field $\psi$ coupled to gravity. Such a state is denoted by $|W,
\psi \rangle$ and, again, is simply referred to as the weave. As
before, it will be characterized by the weave scale ${\mathcal L}$
and the hamiltonian operators $\hat H_{\psi}$ are expected to
fulfill a relation analog to (\ref{eq3: weave-g}), that is, we
shall be able to define an effective hamiltonian $H_{\psi}$ such
that
\begin{eqnarray}
H_{\psi}= \langle W,\psi| \hat H_{\psi} |W,\psi \rangle.
\end{eqnarray}
An approach to this task has been performed by J. Alfaro {\it et
al.} \cite{AMU} for 1/2-spin fermions and the electromagnetic
field. In this approach the effects of the loop structure of space
at the Planck level are treated semiclassically through a
coarse-grained approximation \cite{Gambini Pullin}. This method
leads to the natural appearance of LIV's in the equations of
motion derived from the effective hamiltonian. The key feature
here is that the effective hamiltonian is constructed from
expectation values of dynamical quantities from both the matter
fields and the gravitational field. In this way, when a flat weave
is considered, the expectation values of the gravitational part
will appear in the equations of motion for the matter fields in
the form of coefficients with dependence in both scales,
${\mathcal L}$ and $l_{p}$. When a flat geometry is considered,
the expectation values can be interpreted as vacuum expectation
values for the considered matter fields.

A significant discussion is whether the Lorentz symmetry is
present in the full LQG theory or not \cite{Thiemann1}. For the
present work, we shall assume that Lorentz symmetry is indeed
present in the full LQG theory. This assumption, jointly with the
consideration that the new corrective coefficients are vacuum
expectation values, leads us to consider that the Lorentz symmetry
is spontaneously broken in the effective theory level.

Specially important in  the development of the effective equations
of motion for matter fields is that they are only valid in a
homogenous and isotropic system. From the point of view of a
spontaneous symmetry breakup such a system is unique and,
therefore, a privileged reference frame). It is possible then to
put the equations of motion (and therefore the dispersion
relations) in a covariant form through the introduction of a
four-velocity vector explicitly denoting the existence of a
preferred system. From the cosmological point of view, such a
privileged system does exist, and corresponds to the CMBR
co-moving reference system. For that reason, we shall assume that
the preferred system denoted by the presence of LIV's is the same
CMBR co-moving reference frame, and will use it as the laboratory
system.

In what follows we briefly summarize the obtained dispersion
relations for both, fermions and bosons.

\subsection{Modified Dispersion Relations} \label{fermions}

The dispersion relation for fermions is found to be \cite{AMU}:
\begin{eqnarray} \label{eq3: E for fermions}
E^{2}_{\pm} = p^{2} + 2 \alpha p^{2} + \eta p^{4} \pm 2 \lambda p
+ m^{2},
\end{eqnarray}
where the $\pm$ signs correspond to the helicity state of the
described particle (note that these signs are produced by parity
violation terms in the equations of motion). Additionally, we have
defined the parameters $\alpha$, $\eta$ and $\lambda$ in such a
way that they depend on the scales ${\mathcal L}$ and $l_{p}$ in
the following way:
\begin{eqnarray}
\alpha = \kappa_{\alpha} (l_{p} / {\mathcal L})^{2}, \qquad \qquad
\eta = \kappa_{\eta} \, l_{p}^{2}, \qquad \qquad \mathrm{and}
\qquad \qquad \lambda = \kappa_{\lambda} \, l_{p} / 2 {\mathcal
L}^{2},
\end{eqnarray}
where $\kappa_{\alpha}$, $\kappa_{\eta}$ and $\kappa_{\lambda}$
are adimensional parameters of order 1.

In the case of bosonic particles the dispersion relation consists
of
\begin{eqnarray} \label{eq3: E for bosons}
E^{2} = p^{2} + 2 \alpha p^{2} + \eta p^{4} + m^{2},
\end{eqnarray}
where again we have defined parameters $\alpha$ and $\eta$ as:
\begin{eqnarray}
\alpha = \kappa_{\alpha} (l_{p} / {\mathcal L})^{2}, \qquad \qquad
\mathrm{and} \qquad \qquad \eta = \kappa_{\eta} \, l_{p}^{2}
\end{eqnarray}
where $\kappa_{\alpha}$, $\kappa_{\eta}$ and $\kappa_{\lambda}$
are adimensional parameters of order 1.

As shown in detail in \cite{Alfaro & Palma 2}, only the $\alpha$
correction will be relevant in the computation of the spectrum.
Therefore, in the rest of this work we shall only consider the
following dispersion relation for both, fermions and bosons:
\begin{eqnarray} \label{eq3: E for both}
E^{2} = p^{2} + 2 \alpha p^{2} + m^{2}.
\end{eqnarray}
To conclude, let us mention that the dispersion relation
(\ref{eq3: E for both}) will be used for the physical description
of electrons, protons, neutrons, $\Delta$ and $N$ baryonic
resonances, as well as for mesons $\pi$, $\rho$ and $\omega$.


\section{Threshold Conditions} \label{Trh}

A useful discussion around the effects that LIV's can have on the
propagation of UHECR can be raised through the study of the
threshold conditions for the reactions to take place \cite{Coleman
& Glashow}. To simplify our subsequent discussions, let us use the
following notation for the modified dispersion relations
\begin{eqnarray}
E^{2} = p^{2} + f(p) + m^{2},
\end{eqnarray}
where $f(p)$ is the deformation function of the momentum $p$.

A decay reaction is kinematically allowed when, for a given value
of the total momentum $\vec p_{0} = \sum_{\mathrm{initial}} \vec p
= \sum_{\mathrm{final}} \vec p$, one can find a total energy value
$E_{0}$ such that $E_{0} \geq E_{\mathrm{min}}$. Here
$E_{\mathrm{min}}$ is the minimum value attainable by the total
energy of the decaying products for a given total momentum $\vec
p_{0}$. To find $E_{\mathrm{min}}$, it is enough to take the
individual decay product momenta to be collinear with respect to
the total momentum $\vec p_{0}$ and with the same direction. To
see this, we can variate $E_{0}$ with the appropriate restrictions
\begin{eqnarray}
E_{0} = \sum_{i} E_{i} (p_{i}) + \xi_{j} (p_{0}^{\, j} - \sum_{i}
p_{i}^{\, j}),
\end{eqnarray}
where $\xi_{j}$ are Lagrange multipliers, the $i$ index specifies
the $i$th particle and the $j$ index the $j$th vectorial component
of the different quantities. Doing the variation, we obtain
\begin{eqnarray}
\frac{\partial E_{i}}{\partial p_{i}^{j}} \equiv v_{i}^{j} =
\xi_{j}.
\end{eqnarray}
That is to say, the velocities of all the final produced particles
must be equal to $\xi$. Since the dispersion relations that we are
treating are monotonously increasing in the range of momenta $p >
\lambda$, the momenta can be taken as being collinear and with the
same direction of the initial quantity $\vec p_{0}$.

In this work, we will focus on those cases in which two particles
(say $a$ and $b$) collide to subsequently decay in the
aforementioned final states. For the present discussion, particles
$a$ and $b$ have momenta $\vec p_{a}$ and $\vec p_{b}$
respectively, and the total momentum of the system is $\vec
p_{0}$. It is easy to see from the dispersion relations that we
are considering, that the total energy of the system will depend
only on $p_{a} = |\vec p_{a}|$ and $p_{b} = |\vec p_{b}|$.
Therefore, to obtain the threshold condition for the mentioned
kind of process, we must find the maximum possible total energy
$E_{\mathrm{max}}$ of the initial configuration, given the
knowledge of $p_{a}$ and $p_{b}$. To accomplish this, let us fix
$\vec p_{a}$ and variate the incoming direction of $\vec p_{b} =
\hat n p_{b}$ in
\begin{eqnarray} \label{eq: restricion 2 over E}
E_{0} & = & E_{a}(\vec p_{0} - p_{b} \hat n) + E_{b} (p_{b}) +
\chi (\hat n ^{2} - 1).
\end{eqnarray}
Varying (\ref{eq: restricion 2 over E}) with respect to $\hat n$
($\chi$ is a Lagrange multiplier), we find
\begin{eqnarray}
\hat n ^{i} = v_{a}^{i} p_{b} / 2 \chi.
\end{eqnarray}
In this way we obtain two extremal situations $\chi = \pm v_{a}
p_{b}/2$, or simply $\hat n ^{i} = \pm v_{a}^{i}/v_{a}$. A simple
inspection shows that for the dispersion relations that we are
considering, the maximum energy is given by $\hat n ^{i} = -
v_{a}^{i}/v_{a}$, or in other words, when a frontal collision
takes place.

Summarizing the threshold condition for a two particle ($a$ and
$b$) collision and subsequent decay, can be expressed through the
following requirements:
\begin{eqnarray}
E_{a} + E_{b} \geq \sum_{\mathrm{final}} E_{f},
\end{eqnarray}
with all final particles having the same velocity ($v_{i} = v_{j}$
for any to final particles $i$ and $j$), and
\begin{eqnarray}
p_{a} - p_{b}  = \sum_{\mathrm{final}} p_{f},
\end{eqnarray}
where the sign of the momenta $\sum_{\mathrm{final}} p_{f}$ is
given by the direction of the highest momentum magnitude of the
initial particles.

Our interest in the next subsections is the study of the reactions
involved in high energy cosmic ray phenomena through the threshold
conditions. To accomplish this goal through simple expressions
that are easy-to-manipulate, we shall further use, for the equal
velocities condition, the simplification
\begin{eqnarray}
E_{b}  m_{a} = E_{a}  m_{b},
\end{eqnarray}
valid for the study of parameters coming from the region $f(p) \ll
m^{2}$. This simplification will allow the achievement of bounds
over the order of magnitude of the different parameters involved
in the modified dispersion relations, which are precisely our main
concern.

In the following subsections we study the kinematical effects of
LIV's through the threshold conditions for the reactions involved
in the propagation of UHECR. Since, in this phenomena, photons are
present in the form of low energy particles (the soft photons of
the CMBR), the LQG corrections in the electromagnetic sector of
the theory can be ignored. LQG corrections to the electromagnetic
sector, however, have already been studied for other high energy
reactions such as the Mkn 501 $\gamma$-rays \cite{Alfaro & Palma
1}. Finally, let us recall that in the complete treatment of
threshold conditions, it is possible to learn that only the
$\alpha$ coefficients will be significant \cite{Alfaro & Palma 2}.

\subsection{Photo-Pion Production $\gamma + p \rightarrow p + \pi$}

Let us begin with the photo-pion production $\gamma + p
\rightarrow p + \pi$. Considering the corrections provided in the
dispersion relation (\ref{eq3: E for both}) for fermions and
bosons, we note that, for the photo-pion production to proceed,
the following condition must be satisfied
\begin{eqnarray} \label{eq: pion}
2 \, \delta \alpha \, E_{\pi}^{2} + 4E_{\pi} \omega \geq
\frac{m_{\pi}^{2}(2 m_{p}+m_{\pi})}{m_{p}+m_{\pi}}.
\end{eqnarray}
where $E_{\pi}$ is the produced pion energy and $\delta \alpha =
\alpha_{p} - \alpha_{\pi}$.

\subsection{Resonant Production $\gamma + p \rightarrow \Delta$}

The main channel involved in the photo-pion production is the
resonant production of the $\Delta (1232)$. It can be shown that
the threshold condition for the resonant $\Delta (1232)$ decay
reaction to occur, is
\begin{eqnarray} \label{Delta}
2 \, \delta \alpha \, E^{2} + 4 \omega E \geq m_{\Delta}^{2} -
m_{p}^{2} \, ,
\end{eqnarray}
where $E$ is the incident proton energy and $\delta \alpha =
\alpha_{p} - \alpha_{\Delta}$.

\subsection{Pair Creation $\gamma + p \rightarrow p + e^{+} + e^{-}$}

Pair creation, $\gamma + p \rightarrow p + e^{+} + e^{-}$, is
greatly abundant in the sector previous to the GZK limit. When the
dispersion relations for fermions are considered for both protons
and electrons, it is possible to find
\begin{eqnarray} \label{pair creation}
\delta \alpha \frac{m_{e}}{m_{p}} E^{2} + E \omega \geq m_{e}
(m_{p} + m_{e}), \label{eq: pair}
\end{eqnarray}
where $E$ is the incident proton energy and $\delta \alpha =
\alpha_{p} - \alpha_{e}$.

\subsection{Bounds}

In order to study the threshold conditions (\ref{eq: pion}),
(\ref{Delta}) and (\ref{eq: pair}), in the context of the GZK
anomaly, we must establish some criteria. Firstly, as we have seen
in section \ref{UHECR}, the conventionally obtained theoretical
spectrum provides a very good description of the phenomena up to
an energy $\sim 4 \times 10^{19}$ eV. The main reaction taking
place in this well described region is the pair creation $\gamma +
p \rightarrow p + e^{+} + e^{-}$ and, therefore, no modifications
are present for this reaction up to $\sim 4 \times 10^{19}$ eV. As
a consequence, and since threshold conditions offer a measure of
how modified kinematics is, we will require that the threshold
condition (\ref{eq: pair}) for pair creation not be substantially
altered by the new corrective terms.

Secondly, we have the GZK anomaly itself, which we are committed
to explain. Since for energies greater than $\sim 8 \times
10^{19}$ eV the conventional theoretical spectrum does not fit the
experimental data well, we shall require that LQG corrections be
able to offer a violation of the GZK-cutoff. The dominant reaction
in the violated $E > 8 \times 10^{19}$ region is the photo-pion
production and, therefore, we further require that the new
corrective terms present in the kinematical calculations be able
to shift the threshold significantly to preclude the reaction.

We begin our analysis with the threshold condition for pair
production. In this case we have:
\begin{eqnarray}
\delta \alpha \, \frac{m_{e}}{m_{p}} E^{2} + E \omega \geq m_{e}
(m_{p} + m_{e}),
\end{eqnarray}
with $\delta \alpha = \alpha_{p} - \alpha_{e}$. As is clear from
the above condition, the minimum soft-photon energy
$\omega_{\mathrm{min}}$ for the pair production to occur, is
\begin{eqnarray}
\omega_{\mathrm{min}} = \frac{m_{e}}{E} (m_{p} + m_{e}) - \delta
\alpha \, \frac{m_{e}}{m_{p}} E.
\end{eqnarray}
It follows therefore that the condition for a significant increase
or decrease in the threshold energy for pair production becomes
$|\delta \alpha| \geq m_{p} (m_{p} + m_{e}) / E^{2}$. In this way,
if we do not want kinematics to be modified up to a reference
energy $E_{\mathrm{ref}} = 3 \times 10^{19}$, we must impose the
following constraint
\begin{eqnarray} \label{eq: bound alpha}
|\alpha_{p} - \alpha_{e}| < \frac{(m_{p} + m_{e} ) m_{p} }{
E_{\mathrm{ref}}^{2}} = 9.8 \times 10^{-22}.
\end{eqnarray}
Similar treatments can be found for the analysis of other
astrophysical signals like the Mkn 501 $\gamma$-rays \cite{Stecker
& Glashow}, when the absence of anomalies is considered.

Let us now consider the threshold condition for the photo-pion
production. Taking only the $\alpha$ correction, we have
\begin{eqnarray} \label{eq: pion2}
2 \, \delta \alpha \, E_{\pi}^{2} + 4E_{\pi} \omega \geq
\frac{m_{\pi}^{2}(2 m_{p}+m_{\pi})}{m_{p}+m_{\pi}}.
\end{eqnarray}
It is possible to find that for the above condition to be violated
for all energies $E_{\pi}$ of the emerging pion, and therefore no
reaction to take place, the following inequality must hold
\begin{eqnarray} \label{eq: pi-p}
\alpha_{\pi} - \alpha_{p} > \frac{2 \omega^{2} (m_{p} +
m_{\pi})}{m_{\pi}^{2} (2 m_{p} + m_{\pi})} = 3.3 \times 10^{-24}
\left[ \omega / \omega_{0} \right] ^{2}.
\end{eqnarray}
where $\omega_{0} = K T = 2.35 \times 10^{-4}$ eV is the thermal
CMBR energy. If we repeat these steps for the $\Delta(1232)$
resonant decay, we obtain the following condition
\begin{eqnarray}
\alpha_{\Delta} - \alpha_{p} >  \frac{2 \omega^{2}}{m_{\Delta}^{2}
- m_{p}^{2}} = 1.7 \times 10^{-25} \left[ \omega / \omega_{0}
\right] ^{2}.
\end{eqnarray}

To estimate a range for the weave scale ${\mathcal L}$, let us use
as a reference energy $\omega_{\mathrm{ref}} =
\omega_{\mathrm{min}}$, where $\omega_{\mathrm{min}}$ is the
minimum energy for the reaction to take place, in inequality
(\ref{eq: pion2}), when the condition for a significant increase
in the threshold condition is taken into account (for a primordial
proton reference energy $E_{\mathrm{ref}} = 2 \times 10^{20}$,
this is $\omega_{\mathrm{min}} \sim 2.9 \times \omega_{0}$), and
join the results deduced from the mentioned requirements. Assuming
that the $\kappa_{\alpha}$ parameters are of order 1, as well as
the difference between them for different particles, we can
estimate
---for the weave scale ${\mathcal L}$--- the preferred range
\begin{eqnarray} \label{eq: restriction}
2.6 \times 10^{-18} \,\,\, \mathrm{eV}^{-1} \leq {\mathcal L} \leq
1.6 \times 10^{-17} \,\,\, \mathrm{eV}^{-1},
\end{eqnarray}
where the lefthand and righthand sides come from bounds (\ref{eq:
bound alpha}) and (\ref{eq: pi-p}) respectively (since the
$\Delta$(1232) is just one channel of the photo-pion production,
we shall not consider it to set any bound).

If no GZK anomaly is confirmed in future experimental
observations, then we should state a stronger bound for the
difference $\alpha_{\pi} - \alpha_{p}$. Using the same assumptions
to set the restriction (\ref{eq: bound alpha}) when the primordial
proton reference energy is  $E_{\mathrm{ref}} = 2 \times 10^{20}$
eV, it is possible to find
\begin{eqnarray}
|\alpha_{\pi} - \alpha_{p}| < 2.3 \times 10^{-23}.
\end{eqnarray}
In terms of the length scale ${\mathcal L}$, this last bound may
be read as
\begin{eqnarray}
{\mathcal L} \geq 1.7 \times 10^{-17}  \,\, \mathrm{eV}^{-1},
\end{eqnarray}
which is a stronger bound over ${\mathcal L}$ than (\ref{eq: bound
alpha}), offered by pair creation.


\section{Modified Spectrum} \label{MOD-SPEC}

In this section we show the way in which the LQG correction
$\alpha$ affects the prediction for the theoretical cosmic ray
spectrum. Our approach will be centered on the supposition that
the LQG corrections to the main quantities for the calculation
---such as cross sections and inelasticities of processes--- are,
in a first instance, kinematical corrections, and that the Lorentz
symmetry is spontaneously broken. These assumptions will allow us
to introduce the adequate corrections when a modified dispersion
relation is known.

\subsection{Kinematics}

Having a spontaneous Lorentz symmetry breaking we can use the
still valid Lorentz transformations to express physical quantities
observed in one reference system, in another one. This is possible
since, under a spontaneous symmetry breaking, the group
representations of the broken group preserves its transformation
properties. In particular, it will be possible to relate the
observed 4-momenta in different reference systems through the
usual rule
\begin{eqnarray} \label{eq4: E'-E,p}
p_{\mu}' = \Lambda_{\mu}^{\,\,\, \nu} p_{\nu},
\end{eqnarray}
where $p_{\mu} = (-E, \vec{p})$ is an arbitrary 4-momentum
expressed in a given reference system ${\mathcal K}$, $p_{\mu}'$
is the same vector expressed in another given system ${\mathcal
K}'$, and $\Lambda_{\mu}^{\,\,\, \nu}$ is the usual Lorentz
transformation connecting both systems. Such a transformation will
keep invariant the scalar product
\begin{eqnarray} \label{eq4: prod p-p}
p^{\mu} p_{\mu} = - E^{2} + p^{2},
\end{eqnarray}
as well as any other product. Let us illustrate, for
transformation (\ref{eq4: E'-E,p}), the situation in which
${\mathcal K}'$ is a reference system with the same orientation of
${\mathcal K}$ and which represents an observer with velocity
$\vec \beta$ with respect to ${\mathcal K}$. In this case
$\Lambda_{\mu}^{\,\,\, \nu}$ shall correspond to a boost in the
$\hat \beta = \vec \beta /|\beta|$ direction, and expression
(\ref{eq4: E'-E,p}) will be reduced to
\begin{eqnarray}
E' = \gamma (E - \vec \beta \cdot \vec p), \qquad \mathrm{and}
\qquad \vec p \,\, ' = \gamma (\vec p - \vec \beta E), \label{eq4:
p prima}
\end{eqnarray}
where $\gamma = (1 - \beta^{2})^{-1/2}$. A particular case of this
transformation will be that in which $\vec \beta$ has the same
direction as $\vec p$, and ${\mathcal K}'$ corresponds to the c.m.
reference system, that is to say, the system in which $\vec p \,\,
' = 0$. In such a case we will have $\vec \beta = \vec p /E$ and
$\gamma = E/(E^{2} - p^{2})^{-1/2}$, jointly with the relation
\begin{eqnarray} \label{eq4: E' - gamma}
E' = E/\gamma = (E^{2} - p^{2})^{1/2}.
\end{eqnarray}
In other words, the c.m. energy of a particle with energy $E$ and
momentum $\vec p$ in ${\mathcal K}$ will correspond to the
invariant $(E^{2} - p^{2})^{1/2}$. Furthermore, such energy is the
minimum measurable energy by an arbitrary observer; this can be
confirmed by solving equation $\partial E' / \partial \beta = 0$
from the relation (\ref{eq4: p prima}) and by verifying that the
solution is $\beta = p/E$. This allows us to interpret $E' =
(E^{2} - p^{2})^{1/2}$ as the rest energy of the given particle.
To simplify the notation and the ensuing discussions, let us
introduce the variable $s = (E')^{2}$, where $E'$ is given by
(\ref{eq4: E' - gamma}).

So far in our analysis, relativistic kinematics has not been
modified. Nevertheless, a difference with the conventional
kinematical frame is that in the present theory the product
(\ref{eq4: prod p-p}) will not be independent of the particle's
energy; conversely, we will have the general expression
\begin{eqnarray} \label{eq4: pp - f}
p^{\mu} p_{\mu} =  - f_{a}(E,\vec{p}) - m_{a}^{2},
\end{eqnarray}
where $f_{a}(E,\vec{p})$ is a function of the energy and the
momentum, that represents the LIV provided by the LQG effective
theories. Let us note that expression (\ref{eq4: pp - f}) is just
the modified dispersion relation
\begin{eqnarray} \label{eq4: E - p f}
E^{2} =  p^{2} + f_{a}(E,\vec{p}) + m_{a}^{2}.
\end{eqnarray}
To be consistent $f_{a}(E,\vec{p})$ must be invariant under
Lorentz transformations and, therefore, can be written as a scalar
function of the energy and the momentum.

We have already made mention of the fact that LIV's inevitably
introduce the appearance of a privileged system; in the present
discussion we will choose as such a system the isotropic system
(which by assumption is the co-moving CMBR system), and will
express $f_{a}(E,\vec{p})$ in terms of $E$ and $\vec{p}$ measured
in that system. As may be expected in this situation, $f_{a}$ will
be a function only of the energy $E$ and the momentum norm
$p=|\vec{p}|$, since no trace of a vectorial field could be
allowed when isotropy is imposed. For example, in the particular
case of the dispersion relation for a fermion, the function
$f_{a}(E,\vec{p})$ depends uniquely on the momentum, and can be
written as $f_{a}(p) = 2 \alpha_{a} p^{2}$. For simplicity, we
shall continue using $f_{a}(p)$ instead of $f_{a}(E,\vec{p})$.

Through the recently introduced notation and the use of expression
(\ref{eq4: E' - gamma}), the c.m. energy of an $a$ particle with
mass $m_{a}$ and deformation $f_{a}(p)$ will be
\begin{eqnarray} \label{eq4: s - f}
s_{a}^{1/2} = \sqrt{f_{a}(p) + m_{a}^{2}}.
\end{eqnarray}
Of course, the validity of this interpretation will be subordinate
to those cases in which $s_{a} = f_{a}(p) + m_{a}^{2} > 0$, or,
equivalent, to those states with a time-like 4-momentum. In the
converse, particles with energies and corrections such that $s_{a}
= f_{a}(p) + m_{a}^{2} \leq 0$ will be described by light-like
physical states if the equality holds, or space-like physical
states if the inequality holds.

A new effect provided by LIV's is that, if a reference system
where $p=0$ exists, then in that system the particle will not be
generally at rest. To understand this it is sufficient to verify
that in general the velocity follows $v = \partial E/\partial p
\neq p/E$, and therefore does not vanish at $p=0$. Returning to
the equations in (\ref{eq4: p prima}), we can see that when $\beta
= v =
\partial E / \partial p $, the following result is produced
\begin{eqnarray} \label{eq4: v' -- 0}
\frac{\partial E '}{\partial p '} = \gamma_{v} \left(
\frac{\partial E}{\partial p} - v \right) \frac{\partial
p}{\partial p '} = 0,
\end{eqnarray}
where $\gamma_{v} = (1 - v^{2})^{-1/2}$. Result (\ref{eq4: v' --
0}) shows that the velocity of the system where the particle is at
rest is effectively $v$. There emerges, then, an important
distinction between the phase velocity $\beta = p/E$ of the c.m.
system of a particle, and the group velocity $v =
\partial E / \partial p$ of the same particle.

The above results, for a single particle, can be easily
generalized to a system of many particles. For instance, the total
4-momentum of a system of many particles $p_{\mathrm{tot}}^{\,
\mu} = \sum_{i} p_{i}^{\, \mu}$ will transform through the rule
(\ref{eq4: E'-E,p}), and the scalar product $(p^{\, \mu}
p_{\mu})_{\mathrm{tot}}$ will be an invariant under Lorentz
transformations (as well as any other product). As in the case of
individual particles, we define $\sqrt{s}$ as the total rest
energy measured in the system of c.m. That is to say: $s =
E_{\mathrm{tot}}^{2} - p_{\mathrm{tot}}^{2}$. In the case in which
we have a system composed of a proton with energy $E$ and momentum
$p$, and a photon (from the CMBR) with energy $\omega$ and
momentum $k$ (all these quantities are measured in the laboratory
isotropic system ${\mathcal K}$), the $s$ quantity shall acquire
the form
\begin{eqnarray}
s &=& (E + \omega)^{2} - (\vec p + \vec k)^{2} \nonumber \\
  &=& (E' + \omega')^{2} - (\vec p \, ' + \vec k)^{2},
\end{eqnarray}
where the $E'$ and $p'$ quantities are measured in an arbitrary
reference system. In particular we are interested in the system
where the proton momentum $\vec p \,'$ is null; that is to say,
the system in which $E' = \sqrt{s_{p}} \,$. If $\epsilon$ is the
photon energy in such a system, then
\begin{eqnarray}
s &=& (\sqrt{s_{p}} + \epsilon)^{2} - \epsilon^{2} \nonumber \\
  &=& 2 \sqrt{s_{p}} \,\, \epsilon + s_{p} \, ,
\end{eqnarray}
where we have used the dispersion relation $\omega = k$ (or
$\epsilon = k \, '$) for the CMBR photons. These results will
allow us to express the main kinematical quantities in term of
$\epsilon$ and $s_{p} = f_{p}(p) + m_{p}^{2} \,$. For example, the
Lorentz factor that connects the ${\mathcal K}$ system with the
${\mathcal K}'$ system where $p \, ' = 0$, will be
\begin{eqnarray}
\gamma = E/\sqrt{s_{p}} \, .
\end{eqnarray}
Meanwhile, the Lorentz factor connecting ${\mathcal K}$ with the
c.m. system (that in which $\vec p \, ' + \vec k = 0 $), will be
\begin{eqnarray}
\gamma_{c} = \frac{E + \omega}{s_{p} + 2 \sqrt{s_{p}} \,\,
\epsilon} \simeq \frac{E}{s_{p} + 2 \sqrt{s_{p}} \,\, \epsilon}.
\end{eqnarray}

As a last comment, let us note that to the first order in the
expansion of the dispersion relations in terms of the scales
${\mathcal L}$ and $l_{p}$, when we consider high energy processes
such that $p^{2} \gg f(p) + m^{2}$ we can freely interchange the
momentum $p$ by the energy $E$ in the deviation function $f(p)$.
That is to say, we may consider as a valid relation the following
expression
\begin{eqnarray}
E^{2} = p^{2} + f(E) + m^{2},
\end{eqnarray}
where we have made the replacement $f(p) \rightarrow f(E)$. This
procedure will greatly simplify the next discussions.

\subsection{Modified Inelasticity: $p+\gamma \rightarrow p+x$}

Following the same methods of section \ref{UHECR}, let us obtain a
modified inelasticity $K$ for a process of the type $p+\gamma
\rightarrow p+x$, where $x$ is an emitted particle that, in the
present physical problem in which we are interested, can be a
$\pi$, $\rho$ or $\omega$ meson. We note that the dispersion
relation for the emerging proton (after a collision with a photon)
can be written in the form:
\begin{eqnarray} \label{eq5: Kmod: desarrollo}
E_{p}^2-p_{\, p}^2= f_{p}(E_{p})+m_{p}^{2},
\end{eqnarray}
where $E_{p}$ is the final proton energy. Since the left side of
(\ref{eq5: Kmod: desarrollo}) is invariant under Lorentz
transformations, we can write
\begin{eqnarray}
(E_{p}^{*})^2-(p_{\, p}^{*})^2= f_{p}(E_{p})+m_{p}^{2},
\end{eqnarray}
where the * denotes the quantities measured in the c.m. system. On
the other hand, in such a system, the following conservation
relations of energy and momentum are satisfied:
\begin{eqnarray}
E_{p}^*+E_{x}^*=\sqrt{s} \quad \mathrm{and} \quad
(p_{p}^{*})^2=(p_{x}^{*})^2.
\end{eqnarray}
Substituting both quantities in relation (\ref{eq5: Kmod:
desarrollo}), we obtain
\begin{eqnarray} \label{eq5: Kmod mas des.}
2 \sqrt{s} E_{p}^*= s + s_{p}(E_{p})-s_{x}(E_{x}).
\end{eqnarray}
In the same way, we also have the energy conservation relation in
the laboratory system: $E_{p}+E_{x}=E_{\mathrm{tot}}$. Using the
definition for the inelasticity $K_{x} = \Delta E /E$ for a
process, where $\Delta E = E_{i}-E_{f} \simeq E_{\mathrm{tot}} -
E_{f}$, it is possible to rewrite the energy conservation relation
in the laboratory system in terms of $K_{x}$ through expressions
$E_{x} = K_{x} E$, and $E_{p} = (1 - K_{x}) E$, where $E$ is the
initial energy of the initial proton. Having done this, equation
(\ref{eq5: Kmod mas des.}) now acquires the form:
\begin{eqnarray}
2 \sqrt{s} E_{p}^*= s + s_{p}[(1 - K_{x}) E]-s_{x}[K_{x} E].
\end{eqnarray}
To simplify the development of the inelasticity, let us write the
former relation as $E_{p}^{*}=F(E, K_{x})$, where $F = F(E,K_{x})$
is defined through
\begin{eqnarray} \label{eq5: def F}
F = \frac{1}{2\sqrt{s}}( s + s_{p}[(1 - K_{x}) E]  -s_{x}[K_{x}
E]).
\end{eqnarray}
On the other side, the Lorentz transformation rules give us the
relation between the proton energies in the laboratory system and
the c.m. system respectively. This relation is $E_{p}/\gamma_{c} =
E_{p}^{*} + \beta_{c} \sqrt{E_{p}^{*2} - s_{p}(E_{p})} \cos
\theta$. Joining this expression with (\ref{eq5: def F}), it is
possible to find the general equation for $K_{x}$:
\begin{eqnarray} \label{eq5: ec para Kx}
(1-K_{x}) \sqrt{s} =  \Big( F(E,K_{x})
 + \sqrt{F^{2}(E,K_{x}) - s_{p}[(1-K_{x})E]} \cos \theta \Big).
\end{eqnarray}
It should be noted, however, that the solution for $K_{x}$ from
(\ref{eq5: ec para Kx}) will depend in the $\theta$ angle. For
this reason, once this last equation is resolved, it is convenient
to define the total inelasticity $K$ as the average of $K_{x}$
with respect to the $\theta$ angle. That is to say
\begin{eqnarray} \label{eq5: K mod}
K = \frac{1}{\pi} \int_{0}^{\pi} K_{x} \, d \theta.
\end{eqnarray}
It is relevant to mention that now, as opposed to result
(\ref{eq2: K}), the inelasticity $K$ will be a function of both,
the energy $E$ of the initial proton and the energy $\epsilon$ of
the CMBR photon.

\subsection{The $m_{a}^{2} \rightarrow s_{a} = m_{a}^{2} + f_{a}(E)$
Prescription} \label{sec: prescription}

Let us recall our interpretation relative to the fact that
$s_{a}^{1/2} = (f_{a}(E_{a}) + m_{a}^{2})^{1/2}$ can be understood
as the rest energy of a particle $a$, as a function of the energy
$E_{a}$ that it has in the laboratory system ${\mathcal K}$. As we
have already emphasized, such an interpretation will be valid for
particles with time like 4-momenta.

In the reactions given between high energy protons and the photons
of the CMBR, the whole scenario consists of the collision between
two particles, $p$ and $\gamma$, with the subsequent production of
a certain number of final particles. Let us suppose that $a$ is
one of these particles in the final state. The knowledge of the
inelasticity $K$ for the reaction will allow us to estimate the
average energy $\langle E_{a} \rangle$ with which such a particle
emerges (since $K$ provides the average fraction of energy with
which such a particle is produced). That is to say, on average,
the rest energy of the final particle $a$ will be $s_{a}^{1/2} =
[f_{a}(\langle E_{a} \rangle ) + m_{a}^{2}]^{1/2}$. Moreover, the
knowledge of the inelasticity $K$ will allow us to express $s_{a}$
as a function of the energy $E$ of the initial proton:
\begin{eqnarray} \label{eq5: sa - sa(E)}
s_{a} = s_{a}(E).
\end{eqnarray}
Following our previous interpretation, we can view the recently
described process as a reaction between a proton with mass
$s_{p}^{1/2}$, which loses energy emitting particles $a$ with mass
$s_{a}^{1/2}$ calculated in the previous form. This idealized
reasoning gives us a clear prescription to kinematically modify
those dynamical quantities with which we must work and where
energy conservation is involved. This prescription is:
\begin{eqnarray} \label{eq5: prescrip}
m_{a}^{2} \rightarrow s_{a}(E) = f_{a}(E) + m_{a}^{2},
\end{eqnarray}
where we have expressed correction $f_{a}$ as a function of the
initial energy of the incident proton.

Prescription (\ref{eq5: prescrip}) establishes the notion of an
effective mass which is dependent on the initial energetic content
of a reaction. As a consequence, given the explicit knowledge of the
dependence that a cross section has on the masses and energies of
the involved states, to obtain the modified version, it will be
appropriate to use the discussed prescription.

\subsection{Redshift}

Another important problem related to the introduction of LIV's in
the dispersion relations is whether the redshift relation for the
propagation of particles in a FRW universe is modified. This could
be of great relevance because of the large distances involved in
cosmic ray propagation and, therefore, the possible cumulative
effects. In the case of a FRW universe with scale factor $R$, it
is possible to find that the propagation of particles with
momentum $p^{2} = g^{i j} p_{i} p_{j}$ (with $g_{i j}$ the spatial
part of the FRW metric) obeys the following equation
\begin{eqnarray}
\frac{d}{d \tau}(p R) =  0,
\end{eqnarray}
where $\tau$ is a proper parameter describing the path followed by
the particle. This result is regardless of any particular
dispersion relation.

\subsection{Spectrum and Results}

Introducing the above modifications to the different quantities
involved in the propagation of protons (like the cross section
$\sigma$ and inelasticity $K$), we are able to find a modified
version for the UHECR energy loss due to collisions. Since the
only relevant correction for the GZK anomaly is $\alpha$, we
focused our analysis on the particular case $f(p) = 2 \alpha
p^{2}$. To simplify our model we restricted our treatment to the
case $\alpha > 0$ (consistent with the effective mass
interpretation) and used only $\alpha_{m} \neq 0$, where
$\alpha_{m}$ is assumed to have the same  value for mesons $\pi$,
$\rho$ and $\omega$.

\begin{figure}[tp] 
\centerline{\psfig{figure=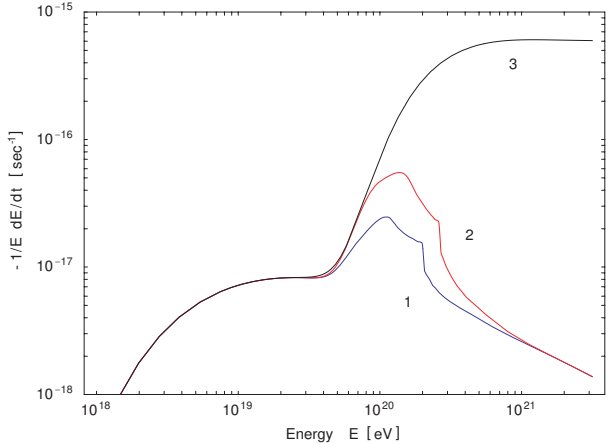,width=8cm}}
\caption{\label{F3} Modified energy loss for UHECR due to
collisions. The figure shows the case $\alpha_{m} \neq 0$, for
three different values of the weave scale ${\mathcal L}$. Curve 1:
$\alpha_{m} = 9 \times 10^{-23}$ (${\mathcal L} \simeq 8.6 \times
10^{-18}$ eV$^{-1}$); curve 2: $\alpha_{m} = 5 \times 10^{-23}$
(${\mathcal L} \simeq 1.2 \times 10^{-17}$ eV$^{-1}$); curve 3:
$\alpha_{m} = 0$ (without modifications).}
\end{figure}

Fig.\ref{F3} shows the modified energy loss $\tau (E)$ for UHECR
obtained for different values of $\alpha_{m}$.
It can be seen therefore how the corrections can affect the main
life time of protons propagating through the CMBR, allowing  a
strong improvement in the distances that protons can reach before
loosing their characteristic energy (for energies greater than $1
\times 10^{20}$ eV). The effects that the LQG corrections
have on the propagation of UHECR are manifest through a decay of
the energy loss in the range $E \sim 1 \times 10^{20}$ eV. To
understand this, recall relation (\ref{eq: pion2}) for the
threshold condition of photo-pion production:
\begin{eqnarray}
2 \, \delta \alpha \, E_{\pi}^{2} + 4E_{\pi} \omega \geq
\frac{m_{\pi}^{2}(2 m_{p}+m_{\pi})}{m_{p}+m_{\pi}}.
\end{eqnarray}
As we saw in section \ref{Trh}, the condition for a significant
increase or decrease in the energy threshold can be calculated as
$|\delta \alpha| \geq (2 m_{p} + m_{\pi})(m_{p} + m_{\pi})/2
E^{2}$. Therefore, for a given value of $\delta \alpha > 0$, the
energy at which the LIV effects start to take place is
\begin{eqnarray}
E^{2} = \frac{1}{2 \delta \alpha} (2 m_{p} + m_{\pi})(m_{p} +
m_{\pi}).
\end{eqnarray}
In the case $\alpha_{m} = 9 \times 10^{-23}$ (curve 1 of
Fig.\ref{F3}), this energy is $E=1.1 \times 10^{20}$ eV, while in
the case $\alpha_{m} = 5 \times 10^{-23}$ (curve 2) this
corresponds to $E=1.5 \times 10^{20}$ eV. Beyond these energy
scales, at about $E \sim 2 \times 10^{20}$ eV, a sharp decay is
observed in the behavior of the curve. This is due to the fact
that the modified inelasticity $K$ will strongly constraint the
energy-momentum phase space accessible to the final states
depending on the initial energy $E$ that the primary proton carries
(recall that now $K$ is a function of the energy $E$ of the
incident proton and the energy $\epsilon$ of the CMBR photon).

\begin{figure}[tp] 
\centerline{\psfig{figure=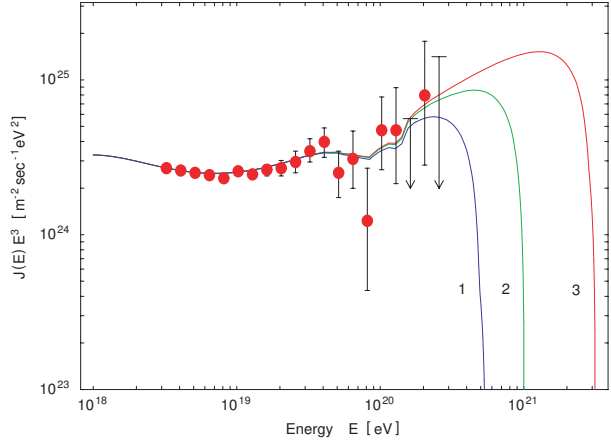,width=8cm}}
\caption{\label{F4} Modified UHECR spectrum and AGASA
observations. The figure shows the modified spectrum $J(E)$
multiplied by $E^{3}$, for uniform distributed sources and without
evolution, for the case $\alpha_{m} = 1.5 \times 10^{-22}$
(${\mathcal L} \simeq 6.7 \times 10^{-18}$ eV$^{-1}$). Three
different maximum generation energies $E_{\mathrm{max}}$ are
shown. These are, curve 1: $5 \times 10^{20}$ eV; curve 2: $1
\times 10^{21}$ eV; and curve 3: $3 \times 10^{21}$ eV.}
\end{figure}

Also, we can find the modified version of the UHECR spectrum for
$\alpha_{m} \neq 0$. Fig.\ref{F4} shows the AGASA observations and
the predicted UHECR spectrum.
The Poisson probabilities of an excess in the five highest energy
bins for the three curves are $P_{1} = 3.6 \times 10^{-4}$, $P_{2}
= 2.6 \times 10^{-4}$ and $P_{3} = 2.3 \times 10^{-4}$. The
Poisson $\chi^{2}$ for the eight highest energy bins are
$\chi^{\,\, 2}_{1} = 10$, $\chi^{\,\, 2}_{2} = 10.9$ and
$\chi^{\,\, 2}_{3} = 11.2$ respectively. The possibility of
reconciling the data with finite maximum generation energies is
significant given that conventional models require infinite
maximum generation energies $E_{\mathrm{max}}$ for the best fit.
 For the lower part of the spectrum (under $E = 4 \times
10^{19}$ eV), the parameters under consideration leave the
spectrum completely unaffected. This is due to the fact that in
such a region the dominant reaction is the pair production, which
has not being modified to obtain the spectrum. A more accurate
study on this issue would require the computation of a modified
inelasticity for the pair creation. Meanwhile, we must content
ourself with the semiqualitative criteria given in section
\ref{Trh} to rule out the parameters.

\section{Conclusions} \label{CONC}

We have seen how the kinematical analysis of the different
reaction taking place in the propagation of ultra high energy
protons can set strong bounds on the parameters to the theory. In
comparison with our previous work, we have eliminated some
previously open possibilities by the particular study of the pair
creation $p + \gamma \rightarrow p + e^{+} + e^{-}$, in the energy
region where this reaction dominates the proton's interactions
with the CMBR. In this way, the only possibility still open is the
correction $\alpha$. If this is the case, a favored region for the
scale length ${\mathcal L}$ estimated through the threshold
analysis would be
\begin{eqnarray}
2.6 \times 10^{-18} \,\,\, \mathrm{eV}^{-1} \leq {\mathcal L} \leq
1.6 \times 10^{-17} \,\,\, \mathrm{eV}^{-1}. \nonumber
\end{eqnarray}
Similarly, the kinematical corrections can be studied in more
detail when their effects are considered in the theoretical
spectrum. In this regard, we have seen how to develop a modified
version of the inelasticity for the photo-pion production, and its
implications in the mean life time of a high energy proton as well
as on the spectrum. To accomplish this last task we have only
assumed a spontaneous Lorentz symmetry break up in the effective
equations of motion, allowing the use of Lorentz transformations
on the dispersion relations. Therefore, result (\ref{eq5: ec para
Kx}) can be used in a more general context than the special case
offered by the LQG framework.

Future experimental developments like the Auger array, the Extreme
Universe Space Observatory (EUSO) and Orbiting Wide-Angle Light
Collectors (OWL) satellite detectors, will increase the precision
and phenomenological description of UHECR. On the more theoretical
side, progress in the direction of a full effective theory, with a
systematic method to compute any correction with a known value for
each coefficient, is one of the next steps in the ``loop''
quantization programme \cite{Thiemann2}. Therefore, it is
important to trace a phenomenological understanding of the
possible effects that could arise as well as the constraints on
LQG, in the high and low energy regimens (for other
phenomenological studies of LQG effects, see for example
\cite{OTHER}).

Recently\cite {collins} the effective theory approach have been under
criticism.

In \cite{alfaro}, the type of terms we used to explain AGASA data are computed
within the Standard model, depending on one arbitrary parameter. The criticism of
\cite{collins} does not apply to this formalism.

\section*{Acknowledgements}
The work of JA is partially supported by Fondecyt 1010967. He
acknowledges the hospitality of LPTENS (Paris); and
financial support from an Ecos(France)-Conicyt(Chile) project. The
work of GAP is partially supported by DAMTP and MIDEPLAN (Chile).
J.A. wants to thank A.A. Andrianov and the organizers of the XVIIIth QFTHEP Workshop
at Saint-Petersburg for their hospitality.


\end{document}